\documentclass[11pt]{article}
\usepackage{epsfig}
\usepackage{graphicx}
\usepackage{amssymb}

\title{Composite Fermions and the Fermion-Chern-Simons Theory}
\author{Bertrand I. Halperin, Physics Department, Harvard University, Cambridge MA 02138}
\begin{document}
\maketitle

ABSTRACT.  The concept of composite fermions, and the related
Fermion-Chern-Simons theory, have been powerful tools for
understanding quantum Hall systems with a partially full lowest Landau
level.  We shall review some of the successes of the
Fermion-Chern-Simons theory, as well as some limitations and
outstanding issues.  

Note: This article is the write-up of an invited talk given at the International
Symposium, 
{\it{Quantum Hall Effect: Past, Present, and Future,}} at
Stuttgart, July 2003.  Proceedings will be published in Physica E.

\bigskip

\section{Introduction}

In the 23 years since Klaus von KIitzing discovered the integer
quantized Hall effect, experiments on quantum Hall systems have
produced a myriad of surprising results.  To understand these, we
have needed to introduce a number of new theoretical tools and
concepts.  The composite fermion picture, and the related
Fermion-Chern-Simons theory have been among the most successful
of these tools, particularly for understanding electrons in the lowest
Landau level, when it is partially full..  The general approach has been
useful for describing both incompressible fractional quantized Hall
states and  compressible unquantized quantum Hall states.

The composite fermion picture was introduced by Jain in 1989, in the
form of trial wavefunctions for the groundstates and quasiparticle
excitations in the principal quantized Hall states, such as 1/3, 2/5, 3/7,
4/9, states which have filling fractions of the form $\nu=p/(2p+1)$,
where $p$ is an integer.\cite{s1,s2}  The composite fermion picture 
correctly
predicted the most prominent observed fractional quantized Hall
states, and Jain showed that the trial wavefunctions had excellent
overlap with the exact wavefunctions for small numbers of particles
on a sphere, for the appropriate filling fractions, with, e.g. Coulomb
interactions between the electrons.

The Fermion-Chern-Simons (FCS) theory was first applied to Jain's
fractional quantized Hall states by Lopez and Fradkin.\cite{s3,s4}   It
was  used by Moore and Read, and Greiter, Wen and
Wilczek,\cite{s5,s6,s7} to gain insight into possible wavefunctions for
observed even-denominator quantized Hall states at $\nu=5/2$ in a
single layer system, and $\nu=1/2$ in certain  bilayer
systems.\cite{x5,x6,x7} The approach was applied to unquantized
fractions, such as $\nu=1/2$ in single layer systems, by Halperin, Lee
and Read \cite{s8}, and by Kalmeyer and Zhang \cite{s9}.  The FCS
theory was particularly useful for understanding dynamic properties of
the quantum Hall systems, such as linear response functions and
transport properties, as well as the dispersion of collective excitations. 
Similar mathematical methods had been used earlier, before the first
applications to quantum Hall systems, by Laughlin and others, in order
to explore properties of systems of ``anyons'', advanced as a model
for high-temperature superconductivity.\cite{s10,s11,s12,s13,s14,s15} 
A Boson-Chern-Simons approach had also been used to develop an
analogy between quantized Hall states and superfluidity, and to
describe the dynamics of quantized Hall states.\cite{s16}

In this article, I shall briefly review the FCS theory and some of its key predictions. I will then say a few word about what is a composite fermion, contrasting the bare FCS particles with the low-energy quasiparticles that emerge from the theory.  I shall also discuss some attempts to formulate a Fermi-liquid like description directly in terms of the low-energy quasiparticles, at compressible fractions such as $\nu=1/2$.  I will not say much about the trial wavefunction approach here, as this will be the subject of Jainendra Jain's presentation.\cite{jt}

\section{Fermion-Chern-Simons theory.}

The FCS approach begins with an exact unitary transformation of the electron problem to a system of ``bare`composite fermions'',  which interact with each other via a fictitious gauge field $\mathbf{a}(\mathbf{r})$, known as the Chern-Simons gauge field, as well as through the usual Coulomb potential.\cite{s8,r1,r2} As a next step, one makes a mean-field approximation on the transformed Hamiltonian.  Finally, one tries to calculate corrections to the mean field results using perturbation theory methods, such as the Random Phase Approximation, Feynman diagram expansions, etc. 

The nature of the Chern-Simons gauge field is that there are an even
integer number $m$ of flux quanta of the Chern-Simons magnetic
field, $b \equiv \nabla \times \mathbf{a}$, 
attached to each fermion.  (By
choosing $m$ to be an even integer, one preserves the antisymmetry
of the wavefunction under interchange of two particles.)  For the
purpose of this presentation, I will confine myself to the case $m=2$.  I
will also assume that the electron spins are completely aligned by the
Zeeman field.

If we make a mean field approximation, (more precisely, a Hartree approximation) to the transformed FCS Hamiltonian, we replace the true Chern-Simons magnetic field $b$, which depends on the position of every particle, by its average value $<b>$.  For a system with a uniform electron density $n_e$, in units where the flux quantum is $2 \pi$, one finds that $<b> = 2 \pi m n_e$ .  In the Hartree approximation, one also replaces the Coulomb interaction of the electrons, with each other and with the positive background,  by its average value, which is just a constant for a uniform system.  Thus the mean field version of the FCS problem describes a collection of non-interacting fermions in a uniform effective magnetic field 
\begin{equation}
B_{\rm{eff}} = B - 2 m \pi n_e \; .
\end{equation}
Recalling that the Landau-level filling factor for electrons is 
defined by $\nu = 2 \pi n_e /B$ , we may define an effective filling 
factor $p$ for the composite fermions by 
$p = 2\pi n_e / B_{\rm{eff}}$.  
It follows that $p^{-1} = \nu^{-1} - m $.  For $m=2$, we then find
that $\nu  = p / (2p+1)$.  If $p$ is an integer, the mean-field theory for
the composite fermions is an integer quantized Hall state with an
energy gap, so we may hope that the interacting electron state will
also have an energy gap and be incompressible.  Of course, the
corresponding values of $\nu$ are precisely the Jain fractions, 
which are the
most prominant fractional quantized Hall states.\cite{s1}

Let us now consider the case where the electron  filling factor is
$\nu=1/2$, so that there are precisely two quanta of magnetic flux
for each electron in the system.  The external magnetic field is then
precisely canceled by the mean Chern-Simons field $<b>$, so that
the effective magnetic field seen by the composite fermions is
$B_{\rm{eff}} = 0$.  The mean-field solution in this case is then a filled
Fermi sea, with a Fermi wavevector $k_F = (  4 \pi n_e)^{1/2} $,
appropriate for a system of spin-aligned fermions in two dimensions. 
This suggests that the groundstate for the interacting electron system
should have  no energy gap, and therefore should be compressible. 
In the presence of scattering due to impurities, we would expect
there to be a finite, non-zero value of the electrical resistivity,
$\rho_{xx}$.  
(We ignore here interaction-corrections to the resistivity, which are
predicted to diverge logarithmically at very low temperatures, but are
negligible in practice.)
Furthermore, if the theory is correct, we should be able
to understand properties of the electron system near $\nu=1/2$ by
using perturbation theory, starting from the Hartree ground state, and
putting in the Coulomb interaction and the fluctuating Chern-Simons
gauge field as perturbations.  There is reason to hope that this
approach will work, even though the mean field solution is not
protected by an energy gap, because the filled Fermi sea has a very
low density of states for multiple particle-hole excitations with low
total energy.  (We contrast this with the original electron problem,
before the unitary transformation, where the partially full Landau level
has an infinitely degenerate ground state before interactions are
turned on.)
On the other hand, there is no small parameter in the perturbation
theory, as the coupling to the fluctuating gauge field is not small.
Hence, quantitative predictions are only possible when one can argue 
that a certain result is correct to all orders in perturbation theory. 

For dynamic calculations, the Hartree approximation is not by itself satisfactory.  However, reasonable results are obtained if one employs the Random Phase Approximation, or a Time-Dependent Hartree Approximation. Here the fermions respond to a self consistent field which includes, in addition to any external driving potentials, the self-consistent Chern-Simons magnetic field $\delta B$ produced by fluctuations in the particle density $<\delta n>$, and a Chern-Simons electric field, given by  
\begin{equation}
\mathbf{e} = - 2 \pi m \hat{z} \times  \mathbf{j} \;  ,
\end {equation}
where $\mathbf{j}$ is the induced current density.

The FCS approach has several strengths and weaknesses, compared to the composite fermion trial wavefunction approach.  A great strength of the FCS approach is that it can address, analytically, the low energy behavior at compressible filling fractions, such as $\nu = 1/2$. It can also address the asymptotic behavior of quantized Hall systems {\emph{close}} to $\nu = 1/2$; for example, it makes non-trivial predictions for the behavior of the energy gaps 
and for the low-energy collective modes in fractional quantized Hall states of the form $\nu=p/(2p+1)$, in the limit $p \to \infty $.\cite{s8,s22,t22a} However, the FCS approach is not very good for calculating the absolute value of energy gaps or of the effective mass for quasiparticles at compressible filling factors.  These are properties which depend on an accurate description of short distance behavior, that are not readily obtained using the FCS perturbation expansion.  As an example, with  the FCS approach, it is not apparent that the energy gaps at fractional quantized Hall states are proportional to the strength of the electron-electron interaction, and must vanish if the interaction goes to zero while the electron mass is held fixed.  In the trial wavefunction approach, this property is guaranteed by projection onto the lowest Landau level,
so that the kinetic energy is automatically a constant, and the differences in energy levels arise only from the interactions.\cite{s1}  Of course, accurate evaluations of energies using trial wavefunctions still require complicated numerical calculations and extrapolations to infinite systems.

\section{Predictions at $\nu=1/2$.}

Let us summarize some of the interesting predictions of the FCS theory for the case of a compressible fraction, such as $\nu = 1/2$, at zero temperature, in the absence of disorder.\cite{s8,r1,r2}  The first prediction is that the system is in fact compressible, i.e., that the energy cost of a small long-wavelength fluctuation in the density is quadratic in  the density fluctuation, and independent of the wavelength, if the singular potential energy due to the long-range Coulomb interaction is subtracted.  (Strictly speaking, the limit should be taken in which the size of the density fluctuation goes to zero while the wavelength is fixed, and then the wavelength goes to infinity.)  

Dynamically, one finds that in the absence of disorder, the longitudinal electrical conductivity vanishes linearly with the wave vector $q$ in the limit $q \to 0$, with a coefficient that can be calculated analytically.  Specifically, at $\nu = 1/2$, for $\mathbf{q}  \parallel \hat{x} $, one finds 
\cite{s8}
\begin{equation}
\label{sigma}
\sigma_{xx}(\mathbf{q}) = \frac{e^2 q}{8 \pi \hbar k_F} \;  .
\end{equation}
 This implies that fluctuations in the electron density relax very slowly at long wavelengths.   In the case of long-range Coulomb interactions, the relaxation rate $\gamma (q)$ is proportional to $q^2$, and is given (in cgs units) by 
 \begin{equation}
 \label{gamma}
 \gamma(q) = \frac{ q^2 e^2}{4 \hbar  k_F  \epsilon } \; ,
 \end{equation}
 where $\epsilon$ is the dielectric constant of the background semiconductor.  In the case of short-range interactions, one finds even slower relaxation, $\gamma(q) \propto q^3$ for $q \to 0$.
 
The longitudinal conductivity $\sigma_{xx}(q)$ is of direct
experimental interest, as it is reflected in the attenuation and velocity
shift in a surface acoustic wave (SAW) experiment. The wavevector
$q$ is the wavevector of the SAW, and we use the zero-frequency
limit of $\sigma_{xx}(q)$, because we are assuming  that the sound
velocity is smaller that the effective Fermi velocity of the composite
fermions.\cite{s8}  This gives an explanation for the anomaly in SAW
propagation at $\nu=1/2$ that was first observed by Willett and
coworkers in 1990.\cite{s17,s18}   For an SAW of sufficiently high
frequency, such that the wavelength is shorter than the
mean-free-path of quasiparticles due to disorder scattering, the
conductivity is enhanced at $\nu=1/2$, proportional to $q$, as
predicted by FCS theory for a clean system.  By contrast, at lower
frequencies, where the SAW wavelength is long compared to the
mean-free-path, one sees the ordinary dc conductivity, which shows
no anomaly a $\nu=1/2$.  

\section{Slightly away from $\nu=1/2$.}

At $\nu=1/2$, the composite fermions see zero effective magnetic 
field, and travel in straight lines until they are scattered by an 
impurity.  Slightly away from $\nu =1/2$, they see a small $B_{\mbox{eff}}$, given by the difference between $B$ and the field corresponding to $\nu=1/2$.  Then the composite fermions will move in circular orbits, with radius\cite{s8}
\begin{equation}
\label{rcstar}
R_c^{*} =  \frac{\hbar k_F}{e |B_{\rm{eff}}|  }  \; .
\end{equation}
The orbit diameter $2R_c^*$ has been measured in geometric resonance experiments with density modulations created by  superimposed periodic structures,\cite{s18a} as well as in SAW experiments,\cite{s19} and in magnetic focusing experiments.\cite{s19a,s19b}   The results agree well with the theoretical prediction, with measured values of $2R_c^*$ which are as large as $\approx1 \mu$m.  This length is of order 100 times larger than the actual cyclotron radius for the electrons in the lowest Landau level.

In Figure 1, we show results of Willett et al.,\cite{s19} comparing measurements of the shift in SAW velocity near $\nu=1/2$, at a frequency of 8.5 GHz, with predictions of the FCS theory of Halperin, Lee and Read. The theoretical curve is broadened by 1.5\% (FWHM) to account for sample inhomogeneities, and there is an adjustment to the overall conductivity scale of the theory.  The theory is essentially an RPA calculation of $\sigma_{xx}(q)$ as a function of magnetic field.  The results are quite different from what one would have obtained if one had ignored the self-consistent Chern-Simons electric field. The minima in the sound velocity occur when $2R_c^*$ is equal to 5/4 times the wavelength of the sound wave, or when the deviation of $B$ from the value at $\nu=1/2$ is equal to $3.83 \hbar c q k_F / e $.  Thus, the fact that the minima coincide in the theoretical and experimental curves is a confirmation that the Fermi momentum of the composite fermions is the same as that for spin-aligned electrons in zero magnetic field.

Experiments which combine SAW propagation and a static density
modulation induced by a superimposed gate array \cite{x20} show
additional peculiar features, which have been explained, at least in
part, using FCS theory.\cite{t20} 

The peculiar wavevector-dependent conductivity (\ref{sigma}) has
implications for experiments in which one tunnels an electron into
the center or the edge of a compressible quantum Hall system, and
also for drag experiments, where one measures the transresistance in
a system of two separated layers, each near $\nu=1/2$.  The FCS
theory has been applied, with some success,  to such
experiments.\cite{t21,t22,t23}

\section{Beyond RPA.}

Other interesting predictions of the FCS theory go beyond the mean
field approximation and the RPA.  For example, it was found by
Halperin, Lee and Read that at $\nu=1/2$ the effective mass of the
quasiparticles should diverge as the energy approaches the Fermi
energy.  For unscreened Coulomb interactions this divergence is only
logartihmic, and is not a large effect in the range accessible to
experiments, but it poses some interesting questions of
principle.\cite{s20,s21}  The divergence in the effective mass does not
affect the long-wavelength linear response functions, as it is
cancelled by interaction effects in the Fermi liquid theory.  Thus the
results of Eqs. (\ref{sigma}) and (\ref{gamma}), which do not involve
the quasiparticle effective mass, are expected to remain correct
beyond the RPA approximation.
The divergent effective mass should be reflected, however, in the energy gaps for the fractional quantized Hall states at $\nu = p/(2p+1)$, in the limit $p \to \infty$.  For the case of unscreened Coulomb interactions, the prediction is \cite{s8,s22}
\begin{equation}
\label{gap}
\Delta_p \sim \frac{\pi e^2}{2 \epsilon l_0} \frac{1}{(2p+1) [\ln (2p+1) + C^{\prime}] } \; ,
\end{equation}
where $l_0$ is the magnetic length, and $C^\prime $ is a constant 
that  depends on the short-range part of the electron-electron 
interaction and cannot be calculated within the FCS theory. 
However, it is believed that the coefficient of the logarithmic term in
the denominator of Eq.~(\ref{gap}) is exact, even though there is 
no small
parameter in the FCS perturbation theory.

In Figure 2, we show results from a recent paper by Morf,
d'Ambrumenil and Das Sarma,\cite{s23} who have performed
calculations of the energy gap for finite systems of up to 18 electrons
on a sphere, and have carefully extrapolated to the limit of an infinite
system, at filling fractions $\nu=1/3, 2/5, 3/7, 4/7$, corresponding to
$p=1,2,3,4$. They find that with a proper choice of $C^{\prime}$, the
gaps fit well to Eq.~(\ref{gap}), and that the fit is better than would
be obtained without the logarithmic correction.  However, it is not
clear {\it a priori} that the asymptotic formula  (\ref{gap}) should
apply for such small values of $p$.  

Unfortunately, it is not possible to draw firm conclusions about the
intrinsic energy gaps from existing experiments.  In order to compare
experiments with theoretical predictions, it has been customary to
subtract a large constant $\Gamma$ from the experimental gaps, in
order to account for effects of impurities.\cite{s24}  This constant
varies from sample to sample, but has been generally taken to be
independent of the $p$, and thus of precise filling fraction.   Recently,
Morf and d'Ambrumenil \cite{s25} have proposed an alternative
correction, with $\Gamma$ varying inversely with $p$, which seems
to give better agreement between theory and experiment. 
However, there is still no satisfactory basic understanding of the effects
of impurities on energy gaps extracted from transport measurements
when the corrections to the gaps are large.

\section{Bare composite fermions and low energy quasiparticles.}

There are some important distinctions between the bare composite fermions which enter the FCS theory, and the low energy quasiparticles which emerge from the theory.  The bare fermions have charge $e$ at all filling fractions.  The low-energy quasiparticles have a reduced charge $e^*$, which in the case of the quantized Hall fractions $\nu=p/(2p+1)$ is given by \cite{laughlin}
\begin{equation}
\label{estar}
e^* = \frac {e}{2p+1} \; .
\end{equation}
At $\nu=1/2$, or the limit $p \to \infty$ this becomes $e^* = 0$, but  
the quasiparticles have an electric dipole moment, $\mathbf{d}$, 
related to the quasiparticle momentum $ \mathbf{k}$,
by \cite{s26}
\begin{equation}
\label{dipole}
\mathbf{d} = e  l_0^2 \hat{z} \times \mathbf{k} \;  .
\end{equation}

It is also known that  the quasiparticles at fractional quantized Hall states are not properly fermions, but rather are anyons, with a statistical angle given by $\theta = \pi (2p-1)/(2p+1)$.\cite{s27} At $\nu=1/2$, however, the statistical angle becomes $\theta = \pi$, corresponding to fermions.   
   
At filling fraction $\nu=p/(2p+1)$, the effective field seen by a composite fermion is related to the external magnetic field by  $B_{\rm{eff}} = B / (2p+1)$.  Thus we see that the cyclotron radius for a quasiparticle of charge $e^*$ given by (\ref{estar}) in the external field $B$ is the same as the effective cyclotron radius given by (\ref{rcstar}), provided one uses the same value of the Fermi momentum $k_F$.

As the low-energy quasiparticles at $\nu=1/2$ are indeed fermions, it should be possible to construct something like a Landau Fermi liquid theory description directly in terms of low-energy neutral, dipolar quasiparticles, as first proposed by Shankar and Murthy.\cite{s28} Various methods for constructing such a description  have been explored by several groups of researchers.\cite{s29,s30,s31,s32}  When done properly, these descriptions all lead to the same predictions as the FCS theory for low energy observable properties, such as the response functions for electrons.  However, the Fermi liquid itself has some rather peculiar properties, as emphasized by Stern et al.\cite{s31}

One peculiar property of the Fermi liquid at $\nu=1/2$ is that energy
is unchanged if a constant $\mathbf{K}$ is added simultaneously to
the momentum of every quasiparticle (i.e., the Fermi surface is
displaced by $\mathbf{K}$).\cite{s30,s31}  Within Landau Fermi liquid
theory this is equivalent to assuming a Landau  interaction parameter 
$F_1 = -1$ for the $l=1$ angular momentum channel.  With
an appropriate 
normalization for $F_l$,  the energy cost of displacing
the Fermi surface by $\mathbf{K}$ is given by
\begin{equation}
\delta E = \frac{n_e K^2}{2 m^*} (1+F_1) \; ,
\end{equation}
where $m^*$ is the quasiparticle effective mass.   

A second peculiarity of the Fermi liquid description is that the local electron density 
$\rho_e(\mathbf{r})$ is related to the {\emph{momentum}} density of quasiparticles $\mathbf{g}(\mathbf{r})$ by \cite{s28}
\begin{equation}
\rho_e= l_0^2 \nabla \times \mathbf{g} \; .
\end{equation}
This relation has a simple interpretation. Comparing it with 
Eq.~(\ref{dipole}), we see that the
electron charge density is given by the divergence of the polarization
density of the dipolar quasiparticles.

The density of quasiparticles is also constrained to be equal to a constant times 
$\nabla \times \mathbf{g} $.  The equations of motion for the quasiparticles are necessarily consistent with this constraint.\cite{s28}

Another way of describing the independence of the energy under a constant momentum displacement is that the neutral quasiparticles have a strong momentum-dependent interaction, which cancels their bare kinetic energy in this situation.  There is no  Chern-Simons vector potential in this Fermi liquid description,  as it has already been eliminated in constructing the neutral quasiparticles. 

Murthy and Shankar have calculated many properties of quantum Hall systems using a Hamiltonian approach for describing composite fermions, which begins with the FCS theory, and proceeds through several additional mathematical transformations and additional approximations.\cite{s29} The objects described by their Hamiltonian, at $\nu=1/2$,  are the neutral quasiparticles  and high energy oscillator modes, describing inter-Landau-level excitations.  Although the correct behavior at low energies is not immediately apparent in the Murthy-Shankar approach, their method does allow them to obtain relatively simple analytic estimates of quantities such as the energy gaps at fractional quantized Hall states, or the energy cost for partial spin polarization of the electrons at $\nu=1/2$, which cannot be obtained from the original FCS theory. These estimates generally agree at the level of $\pm 10\%$ with results from numerical calculations using trial wavefunctions, which are much more difficult to carry out.  The reader is referred to a detailed review article by Murthy and Shankar, which describes their approach.\cite{s29}

\section{Conclusion}

As indicated above, the composite fermion picture and the FCS
theory have been particularly successful in understanding properties
of single layer quantum Hall systems with electrons in the lowest
Landau level.  For electrons above the second Landau level (i.e., for
$\nu > 4$), the composite fermion picture has not, so far, proven
useful.  It appears that the system is better described by the ordinary
Hartree-Fock  theory, where the partially filled Landau level forms
inhomogeneous structures such as charge-density waves or
crystals.\cite{s33,s34,s35}  For electrons in the second Landau level
the system is more delicate.  It appears \cite{s36,s36a} that the 
quantized
Hall states observed at $\nu=5/2$ and $\nu=7/2$  are
well-described by the Moore-Read Pfaffian state,\cite{s5} which can
be understood in the context of FCS theory as a state where a gap is
opened at the Fermi surface of composite fermions due to  a
$p$-wave BCS pairing.\cite{readgreen} 
However, numerical calculations on finite
systems indicate that these states are close to an instability, and small 
changes in the electron-electron interaction at short distances can
drive the system into an anisotropic charge-density-wave state similar
to those in higher Landau levels.\cite{s36,s36a} Experimentally, 
it appears
that application of a magnetic field parallel to the electron layer can
drive this transition.\cite{s37}  Away from $\nu=5/2$ and $7/2$, at 
other filling fractions in the second Landau level,   alternations  have 
been observed in a perpendicular 
magnetic field, between 
fractional quantized Hall states consistent with the composite 
fermion picture, and insulating states suggestive of a straightforward 
Hartree-Fock state.\cite{s37a}

The FCS theory has been applied, with some success to bilayer
systems in the lowest Landau level, but a proper discussion of bilayer
systems is beyond the scope of this article.\cite{s38,s39} Other topics
omitted from this article, where FCS theory has been applied, include
transitions between states with different spin polarizations,\cite{s29}
effects of disorder on transport properties in compressible
states,\cite{s8,s40} and transitions between different fractional quantized
Hall plateaus in the presence of disorder.\cite{s9,s40}

The author has benefited from conversations with many researchers in
the course of his work on quantum Hall systems.  Special thanks are
due to his collaborators in various works most pertinent to this article,
particularly Ady Stern, Steve Simon, Felix von Oppen, Rudolf Morf, 
Nicholas Read,
and Patrick Lee.  Preparation of this manuscript was supported  by
NSF grant   DMR-02-33773.

\bigskip

{\bf{ FIGURE CAPTIONS}}

Figure 1.  Experimental results and theoretical prediction of the shift in surface acoustic wave velocity, as a function of magnetic field, in a GaAs sample with a two-dimensional electron gas below the surface, near filling fraction $\nu=1/2$.  The theoretical curve is the fermion-Chern-Simons prediction of Ref. \cite{s8}, broadened to account for sample inhomogeneity, with an adjustment to the overall scale of the conductivity.  The SAW frequency is 8.5 GHz.  From Ref. \cite{s19}.  

Figure 2.  Energy gaps at quantized Hall fractions of the form $\nu = p/(2p+1)$.  Dots are obtained  exact calculations on systems of up to 18 electrons on a sphere, extrapolated to infinite system size. The dot-dashed curve is a fit to the form expected for non-interacting composite fermions with a constant effective mass; the solid curve includes the logarithmic correction predicted by FCS theory. Each curve has one adjustable constant, chosen to fit the data point at $\nu=1/3$. From Ref. \cite{s23}.

\newpage

\includegraphics[width=\textwidth]{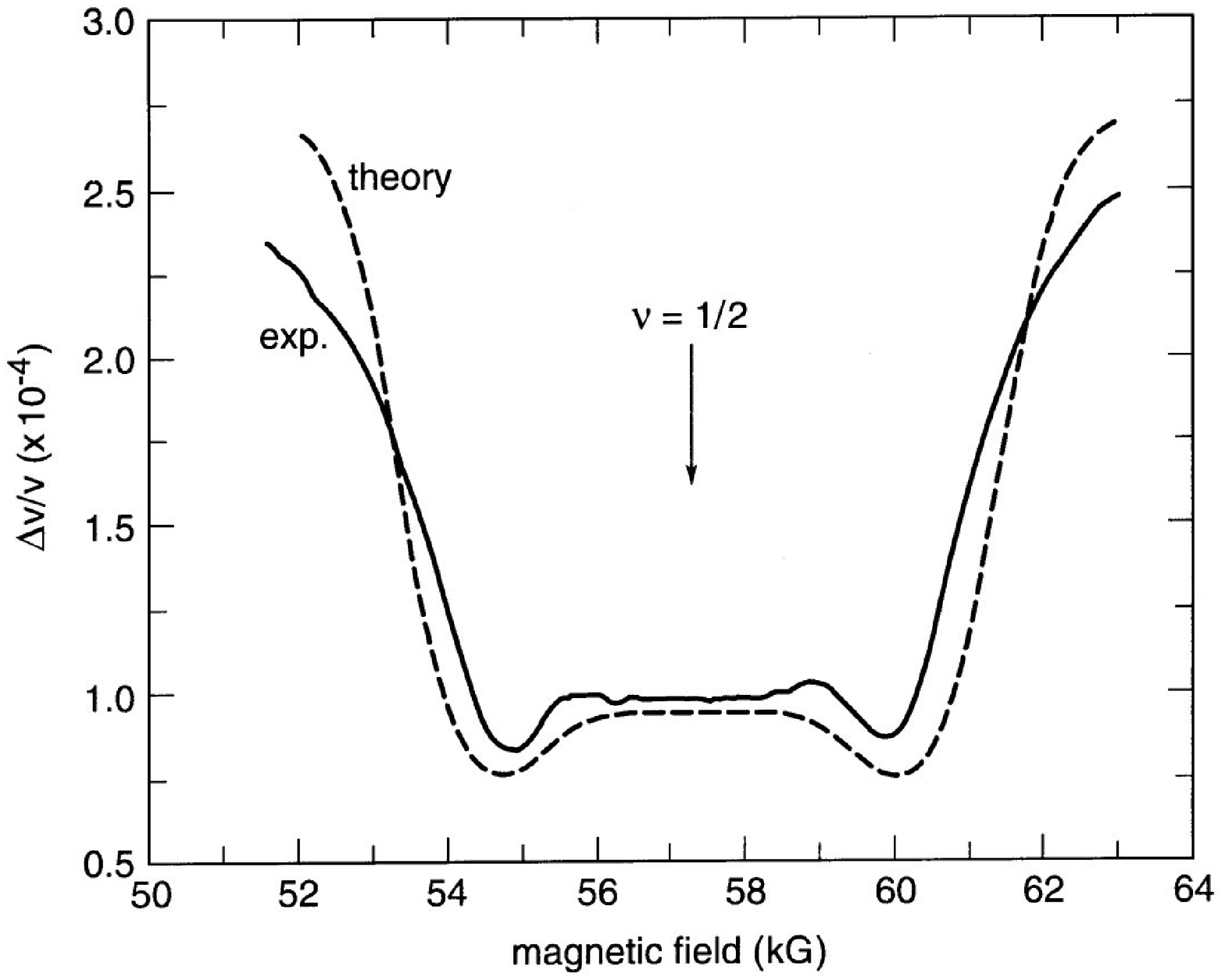}
\vspace{2cm}

{\Large Figure 1. B. I. Halperin}

\newpage

\includegraphics[width=\textwidth]{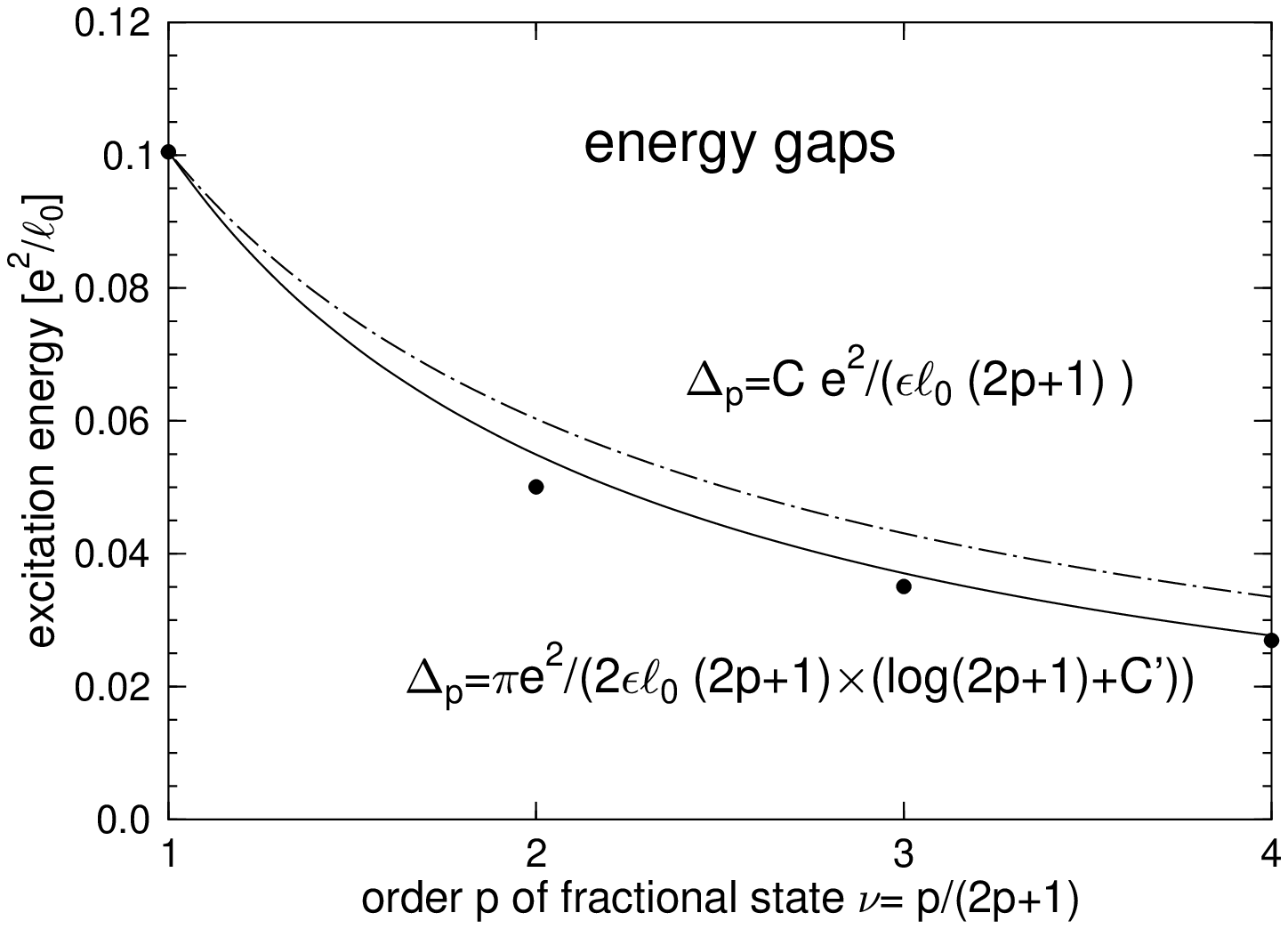}
\vspace{2cm}

{\Large Figure 2. B. I. Halperin}


\begin{thebibliography}{99}

\bibitem{s1} J. K. Jain, Phys. Rev. Lett. {\bf{63}} (1989) 199;
Phys. Rev. {\bf{B40}} (1989) 8079; {\bf{B41}} (1990) 7653.

\bibitem{s2} J.K. Jain, S.A. Kivelson and N. Trivedi,  Phys.
Rev. Lett.{\bf{64}} (1990) 1297.

\bibitem{s3} A. Lopez and E. Fradkin, Phys. Rev. {\bf{B44}}
(1991) 5234.

\bibitem{s4}  A. Lopez and E. Fradkin, Phys. Rev. {\bf{B47}} (1993) 7080.

\bibitem{s5}  G. Moore and N. Read, Nucl. Phys. {\bf{B360}} (1991)
362.


\bibitem{s6} M. Greiter and F. Wilczek, Mod. Phys. {\bf{B4}},
(1990) 1063; Nucl. Phys. B {\bf{370}} (1992) 577.

\bibitem{s7} M. Greiter X.-G. Wen and F. Wilczek, Phys. Rev.
Lett. {\bf{66}} (1991) 3205; Nucl. Phys. {\bf{B374}} (1992) 567.


\bibitem{x5} R.L. Willett, J.P. Eisenstein, H.L. Stormer, D.C.
Tsui, A.C. Gossard and J.H. English, Phys. Rev. Lett. {\bf{59}}
(1987) 1776.

\bibitem{x6} J.P. Eisenstein, G.S. Boebinger, L.N. Pfeiffer,
K.W. West and S. He, Phys. Rev. Lett. {\bf{68}} (1992) 1383.

\bibitem{x7} Y.W. Suen, L.W. Engel, MB. Santos, M. Shayegan and
D.C. Tsui, Phys. Rev. Lett. {\bf{68}} (1992) 1379.

\bibitem{s8}  B.I. Halperin, P.A. Lee and N. Read, Phys. Rev.
{\bf{B47}} (1993) 7312.

\bibitem{s9} V. Kalmeyer and S.-C. Zhang,  Phys. Rev. {\bf{B46}}
(1992) 9889.

\bibitem{s10} R.B. Laughlin, Phys. Rev. Lett. {\bf{60}} (1988)
2677.

\bibitem{s11} A.L. Fetter, C.B. Hanna and R.B. Laughlin, Phys.
Rev. {\bf{B39}} (1989) 9679; C.B. Hanna, R.B. Laughlin and A.L.
Fetter, ibid {\bf{40}} (1989) 8745; {\bf{43}} (1991) 309.

\bibitem{s12} Y.H. Chen, B.I. Halperin, F. Wilczcek and E.
Witten, Int. J. Mod. Phys. {\bf{3}} (1989) 1001.

\bibitem{s13}  B.I. Halperin, J. March-Russell and F. Wilczek,
Phys. Rev. {\bf{B40}} (1989) 8729.

\bibitem{s14} Q. Dai, J.L. Levy, A.L. Fetter, C.B. Hanna and R.B.
Laughlin, Phys. Rev. {\bf{B46}} (1992) 5642.

\bibitem{s15} B. Rejaei and C.W.J. Beenakker, Phys. Rev. {\bf{B43}} 
(1991) 11392.

\bibitem{s16} S.M. Girvin and A.H. MacDonald, Phys. Rev. Lett.
{\bf{58}} (1987) 1252, S.-C. Zhang, H. Hansson and S. Kivelson,
Phys. Rev. Lett. {\bf{62}} (1989) 82; Phys. Rev. Lett. {\bf{62}}
(1989) 980(E); S.-C. Zhang, Int. J. Mod. Phys. {\bf{B6}} (1992)
25; D.-H. Lee and M.P.A. Fisher, Phys. Rev. Lett. {\bf{63}} (1989)
903; N. Read, Phys. Rev. Lett. {\bf{62}} (1989) 86.

\bibitem{jt}  J. Jain, following talk at this symposium. 

\bibitem{r1}  S.H. Simon, in {\it{Composite Fermions}}, edited by O.
Heinonen (World Scientific, Singapore, 1998).

\bibitem{r2}  B.I. Halperin, in {\it{Perspectives in Quantum Hall
Effects}}, edited by S. Das Sarma and A. Pinczuk (Wiley, New York,
1997).

\bibitem{s22} A. Stern and B. I. Halperin, Phys. Rev. B {\bf{52}} (1995) 5890.

\bibitem{t22a}  S.H. Simon and B.I. Halperin, Phys. Rev. {\bf{B48}}
(1993) 17368; S. He, S.H. Simon and B.I. Halperin, Phys. Rev
{\bf{B50}} (1994) 1823.

\bibitem{s17} R.L. Willett, M.A. Paalanen, R.R. Ruel, K.W. West,
L.N. Pfeiffer and D.J. Bishop, Phys. Rev. Lett. {\bf{54}} (1990)
112.

\bibitem{s18}  R.L. Willett, R.R. Ruel, M.A. Paalanen, K.W. West
and L.N. Pfeiffer, Phys. Rev. {\bf{B47}} (1993) 7344.

\bibitem{s18a} W. Kang, H.L. Stormer, L.N. Pfeiffer, K. Baldwin, 
and K.W. West, Phys, Rev, Lett. {\bf{71}} (1993) 3850.

\bibitem{s19} R.L. Willett, R.R. Ruel, K.W. West and L.N.
Pfeiffer, Phys. Rev. Lett. {\bf{71}} (1993) 3846.

\bibitem{s19a} V.J. Goldman, B. Su, and J.K. Jain, Phys. Rev. Lett. 
{\bf{72}} (1994) 2065.

\bibitem{s19b} J.H. Smet {\it{et al}}., Phys. Rev, Lett. {\bf{ 77}} 
 (1996) 2272.

\bibitem{x20} R.L. Willett, K.W. West and L.N. Pfeiffer, Phys.
Rev. Lett. {\bf{78}} (1997) 4478.


\bibitem{t20} F. von Oppen, A. Stern and B.I. Halperin, Phys. Rev.
Lett.  {\bf{80}} (1998) 4494.

\bibitem{t21} S. He, P.M. Platzman, and B.I. Halperin, Phys. Rev. Lett. {\bf{71}} (1993) 777.

\bibitem{t22} See, e.g., L.S. Levitov, A.V. Shytov and B.I.
Halperin, Phys. Rev. {\bf{B64}} (2001) 075322; and references
therein.


\bibitem{t23} See B.N. Narozhny, I.L. Aleiner, and A. Stern, 
Phys. Rev. Lett. {\bf{86}} (2001) 3610; and references therein.

\bibitem{s20}  See Y.B. Kim, P.A. Lee, X.-G. Wen, Phys. Rev.
{\bf{B52}} (1995) 17275; and references therein.

\bibitem{s21} Y.B. Kim and P.A. Lee, Phys. Rev. {\bf{54}} (1996)
2715.


\bibitem{s23} R.H. Morf, N. d'Ambrumenil, and S. Das Sarma, Phys. Rev. B {\bf{66}} (2002) 075408.

\bibitem{s24} R. R. Du, H.L.Stormer, D.C. Tsui, L.N. Pfeiffer, and K. W. West, Phys. Rev. Lett. {\bf{70}} (1993) 2944.

\bibitem{s25} R.H. Morf and N. d'Ambrumenil, Phys. Rev. B (in press), cond-mat/0212304.

\bibitem{laughlin} R. Laughlin, Phys. Rev. Lett. {\bf{50}}  (1983) 1395.

\bibitem{s26} N. Read, Semicond. Sci. Technol. {\bf{9}} (1994) 1859; Surf. Sci, {\bf{361/362}} (1996) 7. 

\bibitem{s27} B. I. Halperin,  Phys.  Rev. Lett.  {\bf{52}}  (1984) 1583, 2390(E). 

\bibitem{s28} R. Shankar and G. Murthy, Phys. Rev. Lett. {\bf{79}} (1997) 4437.

\bibitem{s29} G. Murthy and R. Shankar, Rev. Mod. Phys. (in press).

\bibitem{s30} B.I.Halperin and A. Stern, Phys. Rev. Lett. {\bf{80}} (1998) 5457.

\bibitem{s31} A. Stern, B.I. Halperin, F. von Oppen, and S. H. Simon, Phys. Rev. {\bf{B59}} (1999) 12567.

\bibitem{s32} V. Pasquier and F.D.M. Haldane, Nucl. Phys. B {\bf{516}} (1998) 719; N. Read, Phys. Rev. B {\bf{58}} (1998) 16262; D.-H. Lee, Phys. Rev. Lett. {\bf{80}} (1998) 4745.

\bibitem{s33} M. P. Lilly, et al., Phys. Rev. Lett. {\bf{82}} (1999) 394;  
R.R. Du, et  al., Solid State Commun. { \bf{109}} (1999) 389.

\bibitem{s34} M.M. Fogler, A.A. Koulakov, and B.I. Shklovskii, Phys. Rev. B {\bf{54}} (1996) 1853; A.A. Koulakov, M.M. Fogler, and B.I. Shklovskii, Phys. Rev. Lett. {\bf{76}} (1996) 499; 
R. Moessner and J.T. Chalker, Phys. Rev. B {\bf{54}} (1996) 5006.

\bibitem{s35} For more recent work, see D. Barci and E. Fradkin, Phys.
Rev. B {\bf{65}} (2002) 245320, and references therein.


\bibitem{s36} R. H. Morf, Phys. Rev. Lett. {\bf{80}} (1998) 1505.

\bibitem{s36a} E.H. Rezayi and F.D.M. Haldane, Phys. Rev. Lett. {\bf{84}} (2000) 4685. 

\bibitem{readgreen} N. Read and D. Green, 
Phys. Rev. B {\bf{61}} (2000) 10267.
  
\bibitem{s37} W. Pan, et al., Phys. Rev. Lett. {\bf{83}} (1999) 820; 
M. Lilly, et al., Phys. Rev. Lett. {\bf{83}} (1999) 824.

\bibitem{s37a} J.P. Eisenstein, K.B. Cooper, L.N. Pfeiffer, and 
K.W. West, Phys. Rev. Lett. {\bf{88}} (2002) 076801.


\bibitem{s38} See, e.g., B.I. Halperin, Surf. Sci. {\bf{305}} (1994) 1, and references therein

\bibitem{s39} N.E. Bonesteel, Phys. Rev. {\bf{B48}} (1993) 11484.

\bibitem{s40} D.B, Chklovskii and P.A. Lee, Phys. Rev. B. {\bf{48}} (1993) 18060. 
\end{thebibliography}
 \end{document}